\documentclass[12pt]{article}
\usepackage{amssymb}
\usepackage{amsmath}

\newcommand{\be}{\begin{eqnarray}}
\newcommand{\ee}{\end{eqnarray}}
\newcommand{\non}{\nonumber}
\newcommand{\id}{\mathbb{I}}
\newcommand{\tr}{\mathop{\rm tr}\nolimits}

\begin{document}

\begin{titlepage}
\strut\hfill UMTG--275
\vspace{.5in}
\begin{center}

\LARGE Algebraic Bethe ansatz for singular solutions\\
\vspace{1in}
\large Rafael I. Nepomechie \footnote{nepomechie@physics.miami.edu}
and Chunguang Wang \footnote{c.wang22@umiami.edu}\\[0.8in]
\large Physics Department, P.O. Box 248046, University of Miami\\[0.2in]  
\large Coral Gables, FL 33124 USA\\

\end{center}

\vspace{.5in}

\begin{abstract}
The Bethe equations for the isotropic periodic spin-1/2 Heisenberg
chain with $N$ sites have solutions containing $\pm i/2$ that are
singular: both the corresponding energy and the algebraic Bethe ansatz
vector are divergent.  Such solutions must be carefully regularized.
We consider a regularization involving a parameter that can
be determined using a generalization of the Bethe equations.  These
generalized Bethe equations provide a practical way of determining
which singular solutions correspond to eigenvectors of the model.
\end{abstract}

\end{titlepage}

\setcounter{footnote}{0}

\section{Introduction}\label{sec:intro}

It is well known that the isotropic periodic spin-$\frac{1}{2}$ Heisenberg quantum spin 
chain with $N$ sites, with Hamiltonian
\be
H =  {1\over 4}\sum_{n=1}^{N}  \left( 
\vec \sigma_{n} \cdot \vec \sigma_{n+1} - 1 \right)  \,, \qquad \vec 
\sigma_{N+1} \equiv \vec \sigma_{1} \,,
\label{Heisenberg}
\ee 
can be solved by algebraic Bethe ansatz (ABA):
the eigenvalues are given by
\be
E = - {1\over 2} \sum_{k=1}^{M} \frac{1}{\lambda_{k}^{2} + 
\frac{1}{4}} \,,
\label{energy}
\ee 
and the corresponding $su(2)$ highest-weight eigenvectors are given by the 
Bethe vectors
\be
|\lambda_{1} \,, \ldots \,, \lambda_{M} \rangle =
B(\lambda_{1}) \cdots B(\lambda_{M}) |0\rangle \,,
\label{vec}
\ee
where $|0\rangle$ is the reference state with all spins up,
$\{\lambda_{1} \,, \ldots \,, \lambda_{M}\}$
are distinct and satisfy the Bethe equations 
\be
\left(  \frac{\lambda_{k} + \frac{i}{2}}
{\lambda_{k} - \frac{i}{2}} \right)^{N} 
= \prod_{\scriptstyle{j \ne k}\atop \scriptstyle{j=1}}^M 
\frac{\lambda_{k} - \lambda_{j} + i} 
{\lambda_{k} - \lambda_{j} - i}
\,, \qquad k = 1 \,, \cdots \,, M \,,
\label{BAE}
\ee
and $M = 0, 1, \ldots,  \frac{N}{2}$. The spin $s$ of the state is given 
by $s=\frac{N}{2}-M$.
(See, for example, \cite{Faddeev:1996iy, Korepin:1993}.)

It is also well known that the so-called two-string 
$(\lambda_{1}\,,\lambda_{2}) = (\frac{i}{2}\,, -\frac{i}{2})$ is an
{\it exact} solution of the Bethe equations for $N \ge 4$.
This fact is particularly easy to see from the 
Bethe equations in the pole-free form
\be
(\lambda_{1} + \frac{i}{2})^{N} (\lambda_{1} - \lambda_{2} - i)
&= (\lambda_{1} - \frac{i}{2})^{N} (\lambda_{1} - \lambda_{2} + i) 
\,, \non \\
(\lambda_{2} + \frac{i}{2})^{N} (\lambda_{2} - \lambda_{1} - i)
&= (\lambda_{2} - \frac{i}{2})^{N} (\lambda_{2} - \lambda_{1} + i) \,.
\ee 
This solution is singular, as both the
corresponding energy
(\ref{energy}) and Bethe vector (\ref{vec}) are 
divergent.\footnote{While the divergence of the energy is obvious, 
the divergence of the Bethe vector is a consequence of our 
non-standard conventions, which we specify in Section \ref{sec:analysis} below. In 
the standard conventions, the Bethe vector would instead be null.}
Clearly, it is necessary to regularize this  
solution. The naive regularization
\be
\lambda_{1}^{naive} = \frac{i}{2} + \epsilon\,, \qquad 
\lambda_{2}^{naive} = -\frac{i}{2} + \epsilon 
\label{naive}
\ee
gives the correct value of the energy in the $\epsilon 
\rightarrow 0$ limit, namely, $E=-1$.

What is perhaps not so well known is that this naive regularization gives a 
wrong result for the eigenvector.\footnote{Difficulties with constructing the eigenvector corresponding to 
the Bethe roots $\pm \frac{i}{2}$ were already noted in 
\cite{Siddharthan:1998, Wal:1999}.}
Indeed, the vector 
$ \lim_{\epsilon \rightarrow 0}\ |\lambda_{1}^{naive} \,,  
\lambda_{2}^{naive} \rangle$
is finite, but it is {\it not} an eigenvector of the Hamiltonian! For 
example, in the case $N=4$, we easily find with {\tt Mathematica} 
that
\be 
\lim_{\epsilon \rightarrow 0}\ |\lambda_{1}^{naive} \,,  
\lambda_{2}^{naive} \rangle = 
(0, 0, 0, 2, 0, 0, -2, 0, 0, 0, 0, 0, 2, 0, 0, 0)\,,
\ee
while the correct eigenvector with $E=-1$ and $s=0$ is known to be 
instead \footnote{For any even $N$, the Bethe vector corresponding to the 
2-string $\pm \frac{i}{2}$ can be expressed as \cite{Essler:1991wd}
\be
\sum_{k=1}^{N}(-1)^{k}S_{k}^{-}S_{k+1}^{-} |0\rangle \,.
\ee
One can easily verify that for $N=4$ this vector is indeed proportional to (\ref{correct}).}  
\be
(0, 0, 0, 2, 0, 0, -2, 0, 0, -2, 0, 0, 2, 0, 0, 0) \,.
\label{correct}
\ee

We further observe that, for general values of $N$, the correct eigenvector
can be obtained within the ABA approach by introducing a suitable additional correction of order
$\epsilon^{N}$ to the Bethe roots:\footnote{Such higher-order
corrections of singular Bethe roots were already noted in Eq.  (3.4) of
\cite{Beisert:2003xu} and studied further in \cite{Beisert:2004hm}.}
\be
\lambda_{1} = \frac{i}{2} + \epsilon + c\, \epsilon^{N}\,, \qquad 
\lambda_{2} = -\frac{i}{2} + \epsilon \,,
\label{regularized}
\ee
where the parameter $c$ is independent of $\epsilon$. Returning to 
the example of $N=4$, we find
\be 
\lim_{\epsilon \rightarrow 0}\ |\lambda_{1} \,,  
\lambda_{2} \rangle = 
(0, 0, 0, 2, 0, 0, -2, 0, 0, i c, 0, 0, 2, 0, 0, 0)\,.
\label{limit}
\ee
Comparing with (\ref{correct}), we see that the requisite value of the
parameter in this case is $c=2i$.

In Section \ref{sec:analysis}, we address the question of how to
determine in a systematic way the parameter $c$ in
(\ref{regularized}), which (as we have already seen) is necessary for
obtaining the correct eigenvector.  Clearly, it is not a matter of
simply solving the Bethe equations (\ref{BAE}), since they are not
satisfied by (\ref{regularized}) for $\epsilon$ finite.  Indeed, we
shall find that the Bethe equations themselves acquire
$\epsilon$-dependent corrections.  These ``generalized'' Bethe
equations (see Eq.  (\ref{Fconditions}) below) constitute our main new
result.  In Section \ref{sec:general}, we extend this approach to
general singular solutions, i.e., solutions of the Bethe equations
where two of the roots are $\pm\frac{i}{2}$.  Typically, there are
many such solutions, but relatively few correspond to eigenvectors of
the model.  We find that the generalized Bethe equations provide a
practical way of determining which of the singular solutions
correspond to eigenvectors.  Section \ref{sec:conclusion} summarizes our
main conclusions.

Singular solutions do not appear in a related model, namely, the
Heisenberg chain with twisted boundary conditions.  A small twist
angle $\phi$ then plays a similar role to our parameter $\epsilon$.
This alternative approach for dealing with singular solutions is
briefly considered in appendix B.3 of \cite{Goetze:2010} and in
section 2.1 of \cite{Arutyunov:2012tx}.  Since the twist breaks the
$su(2)$ symmetry, the Bethe vectors are no longer highest-weight
vectors.  Our point of view is that the isotropic periodic Heisenberg
chain for finite $N$ is a well-defined model, and therefore should be
understandable independently of other models; it is only its Bethe
ansatz solution that is not completely well defined.

Yet another approach for constructing the Bethe vectors corresponding 
to singular solutions, involving Sklyanin's separation of variables, 
was carefully analyzed in \cite{Mukhin:2009}.

\section{Determining the parameter}\label{sec:analysis}

We begin by briefly establishing our conventions. Following 
\cite{Faddeev:1996iy}, the $R$-matrix is given by
\be
R_{a_{1} a_{2}}(\lambda) = \lambda \id_{a_{1} a_{2}} + i {\cal P}_{a_{1} a_{2}} \,,
\ee
where $\id$ and ${\cal P}$ are the $4 \times 4$ identity and permutation matrices, 
respectively. However, as explained below, we choose a 
different normalization for the Lax operator, namely,
\be
L_{n a}(\lambda) = \frac{1}{(\lambda + \frac{i}{2})}\left[(\lambda - 
\frac{i}{2})\id_{n a} + i {\cal P}_{n a} \right] \,,
\label{Lax}
\ee
which diverges for $\lambda=-\frac{i}{2}$.
As usual, the monodromy matrix is given by
\be
T_{a}(\lambda) = L_{N a}(\lambda)  \cdots L_{1 a}(\lambda) = \left(
\begin{array}{cc}
    A(\lambda) & B(\lambda) \\
    C(\lambda) & D(\lambda) 
\end{array}    \right) \,,
\ee 
and the transfer matrix is given by
\be
t(\lambda) = \tr_{a} T_{a}(\lambda) = A(\lambda) + D(\lambda) \,.
\ee
The reference state is denoted by $|0\rangle = {1 \choose 0}^{\otimes N}$.

We next recall the action of the transfer matrix on an off-shell 
Bethe vector (\ref{vec}) \cite{Faddeev:1996iy}
\be
t(\lambda) |\lambda_{1} \,, \ldots \,, \lambda_{M} \rangle &=& 
\Lambda(\lambda) |\lambda_{1} \,, \ldots \,, \lambda_{M} \rangle \non 
\\
&+& 
\sum_{k=1}^{M} F_{k}(\lambda,\{\lambda\})
B(\lambda_{1}) \cdots \widehat{B}(\lambda_{k}) \cdots B(\lambda_{M})B(\lambda) |0\rangle \,,
\label{offshell1}
\ee
where a hat is used to denote an operator that is omitted, and
\be
\Lambda(\lambda) &=& 
\prod_{j=1}^{M}\left(\frac{\lambda-\lambda_{j}-i}{\lambda-\lambda_{j}}\right)
+\left(\frac{\lambda-\frac{i}{2}}{\lambda+\frac{i}{2}}\right)^{N}
\prod_{j=1}^{M}\left(\frac{\lambda-\lambda_{j}+i}{\lambda-\lambda_{j}}\right)\,, 
\label{Lambda}\\
F_{k}(\lambda,\{\lambda\}) &=& \frac{i}{\lambda-\lambda_{k}}\left[
\prod_{j\ne k}^{M}\left(\frac{\lambda_{k}-\lambda_{j}-i}{\lambda_{k}-\lambda_{j}}\right)
-\left(\frac{\lambda_{k}-\frac{i}{2}}{\lambda_{k}+\frac{i}{2}}\right)^{N}
\prod_{j\ne k}^{M}\left(\frac{\lambda_{k}-\lambda_{j}+i}{\lambda_{k}-\lambda_{j}}\right) \right]
\,. \label{Fs}
\ee
The Bethe equations (\ref{BAE}) are precisely the conditions 
$F_{k}(\lambda,\{\lambda\})=0$, which ensure that the ``unwanted'' 
terms vanish, in which case the Bethe vector $|\lambda_{1} \,, 
\ldots \,, \lambda_{M} \rangle$ is an eigenvector of the transfer 
matrix $t(\lambda)$, with corresponding eigenvalue $\Lambda(\lambda)$ given by (\ref{Lambda}).
In particular, for $M=2$, the relation (\ref{offshell1}) reduces to
\be
t(\lambda) |\lambda_{1} \,, \lambda_{2} \rangle = 
\Lambda(\lambda) |\lambda_{1} \,, \lambda_{2} \rangle  
+ F_{1}(\lambda,\{\lambda\}) B(\lambda_{2})B(\lambda)|0\rangle 
+ F_{2}(\lambda,\{\lambda\}) B(\lambda_{1})B(\lambda)|0\rangle \,,
\label{offshell2}
\ee
which holds for generic values of $\lambda\,, \lambda_{1}$ and $\lambda_{2}$. 

Let us now focus on the special case of the two-string solution $\pm
\frac{i}{2}$.  As already mentioned in the Introduction, the
corresponding Bethe vector $|\frac{i}{2}, -\frac{i}{2}\rangle$ is 
singular: some of its components have the form $0/0$.  (If we had
defined the Lax operator (\ref{Lax}) without dividing by $(\lambda +
\frac{i}{2})$ as in \cite{Faddeev:1996iy}, then the corresponding
Bethe vector would instead be null \cite{Siddharthan:1998}.)  In particular, the 
creation operator $B(\frac{i}{2})$ is finite, but $B(-\frac{i}{2})$ is 
singular.

Let us first consider the naive regularization 
(\ref{naive}). The key observation is that, for $\epsilon \rightarrow 0$,
the most singular matrix elements of $B(\lambda_{2}^{naive})$ are of 
order $\frac{1}{\epsilon^{N}}$. (See (\ref{Bbehavior}).) It follows from the off-shell relation 
(\ref{offshell2}) that, for $\epsilon \rightarrow 0$, the 
coefficients $F_{1}$ and $F_{2}$ must satisfy
\be
F_{1}(\lambda,\{\lambda\}) \sim \epsilon^{N+1} \,, \qquad 
F_{2}(\lambda,\{\lambda\}) \sim \epsilon \,,
\label{Fconditions}
\ee
in order that the Bethe vector $\lim_{\epsilon \rightarrow 0}\ |\lambda_{1}^{naive} \,,  
\lambda_{2}^{naive} \rangle$
be an eigenvector of the transfer
matrix.  However, explicit computation using (\ref{naive}) shows that
$F_{1}(\lambda,\{\lambda\}) \sim \epsilon^{N}$ (instead of $
\epsilon^{N+1}$) and $F_{2}(\lambda,\{\lambda\}) \sim 1$ (instead of $
\epsilon$).  Hence, the ``unwanted'' terms in (\ref{offshell2}) are finite (do not vanish), which
explains why the corresponding Bethe vector is not an
eigenvector.\footnote{The fact that $B(\lambda_{2}^{naive})$ has
matrix elements of order $\frac{1}{\epsilon^{N}}$ suggests that
$|\lambda_{1}^{naive} \,, \lambda_{2}^{naive} \rangle \sim
\frac{1}{\epsilon^{N}}$.  However, as shown in the Appendix, this
vector is finite for $\epsilon \rightarrow 0$.}

Let us therefore consider the regularization (\ref{regularized}).  The
leading behavior of $B(\lambda_{1})$ and $B(\lambda_{2})$ as $\epsilon
\rightarrow 0$ remains the same as with the naive regularization; i.e., $B(\lambda_{1}) \sim 
\frac{1}{\epsilon^{N}}$ and $B(\lambda_{2}) \sim 1$.
Hence, the conditions
(\ref{Fconditions}) must still be satisfied to ensure that the Bethe 
vector is an eigenvector of the transfer matrix.  Explicit computation
using (\ref{regularized}) gives
\be
F_{1}(\lambda,\{\lambda\}) = \left(\frac{c+2 
i^{-(N+1)}}{\lambda-\frac{i}{2}}\right)\epsilon^{N} + 
O(\epsilon^{N+1}) \,,
\quad 
F_{2}(\lambda,\{\lambda\}) =  \left(\frac{2i -
i^{-N}c}{\lambda+\frac{i}{2}}\right) +  O(\epsilon) \,.
\ee 
For even $N$, both conditions (\ref{Fconditions}) can be 
satisfied by setting
\be
c = 2 i (-1)^{N/2} \,,
\label{param}
\ee 
which reproduces our earlier result for $N=4$ (see below 
Eq. \ref{limit}). We have also 
explicitly verified that, 
for $N=6$, the ABA Bethe vector constructed using (\ref{regularized}) 
and (\ref{param}) is indeed an eigenvector of the 
Hamiltonian.\footnote{It was claimed in \cite{Wal:1999} that the Bethe ansatz 
fails for this case.}
Interestingly,  the two
conditions (\ref{Fconditions}) cannot be simultaneously satisfied for odd 
$N$, implying that the two-string $\pm \frac{i}{2}$ is {\it not} a bona fide 
solution for odd $N$.\footnote{For $N=5$, the 
Clebsch-Gordan theorem implies that there are five 
highest-weight eigenvectors with $s=\frac{1}{2}$; and we have explicitly verified 
that all of these eigenvectors can be constructed with Bethe roots other than $\pm 
\frac{i}{2}$, thereby directly proving that the solution $\pm 
\frac{i}{2}$ must be discarded.}

We note that the regularization (\ref{regularized}) can be slightly 
generalized. 
Indeed, we can introduce a two-parameter regularization
\be
\lambda_{1} = \frac{i}{2} + \epsilon + c_{1}\, \epsilon^{N}\,, \qquad 
\lambda_{2} = -\frac{i}{2} + \epsilon + c_{2}\, \epsilon^{N} \,.
\label{regularized2}
\ee
The conditions (\ref{Fconditions}) now imply (for even $N$) that
\be
c_{1}-c_{2} = 2 i (-1)^{N/2} \,.
\ee 
For finite $\epsilon$, the corresponding energy (\ref{energy}) 
depends only on the difference $c_{1}-c_{2}$.
If we impose the additional constraint $\lambda_{1} = 
\lambda_{2}^{*}$ \cite{Vladimirov:1986}, then we obtain $c_{1}=c_{2}^{*} = i (-1)^{N/2}$. In short, 
for even $N$,
a regularization of the singular solution $\pm \frac{i}{2}$ that 
produces the correct eigenvector in the $\epsilon \rightarrow 0$ limit, 
and also satisfies $\lambda_{1} = \lambda_{2}^{*}$, is given by
\be
\lambda_{1} = \frac{i}{2} + \epsilon + i (-1)^{N/2} \, \epsilon^{N}\,, \qquad 
\lambda_{2} = -\frac{i}{2} + \epsilon - i (-1)^{N/2}\, \epsilon^{N} \,.
\label{regularized3}
\ee

\section{General singular solutions}\label{sec:general}

We now consider a general singular solution of the Bethe equations, 
which has the form 
\be
\{ \frac{i}{2}\,, -\frac{i}{2}\,, \lambda_{3}\,, \ldots \,, 
\lambda_{M} \} \,,
\label{gensing}
\ee 
where $\lambda_{3}\,, \ldots \,, \lambda_{M}$ are distinct and are not equal to $\pm \frac{i}{2}$. 
Proceeding as before, we regularize the first two roots as in Eq. (\ref{regularized}).
The Bethe equations (\ref{BAE}) imply that the last $M-2$ roots $\{  
\lambda_{3}\,, \ldots \,, \lambda_{M} \}$ obey
\be
\left(  \frac{\lambda_{k} + \frac{i}{2}}
{\lambda_{k} - \frac{i}{2}} \right)^{N-1} 
\left(\frac{\lambda_{k} - \frac{3i}{2}}
{\lambda_{k} + \frac{3i}{2}}\right)
=  \prod_{\scriptstyle{j \ne k}\atop \scriptstyle{j=3}}^{M}
\frac{\lambda_{k} - \lambda_{j} + i} 
{\lambda_{k} - \lambda_{j} - i}
\,, \qquad k = 3 \,, \cdots \,, M \,.
\label{BAE2}
\ee

We again impose the two generalized Bethe equations
\be
F_{1}(\lambda,\{\lambda\}) \sim \epsilon^{N+1} \,, \qquad 
F_{2}(\lambda,\{\lambda\}) \sim \epsilon \,,
\label{Fconditionsagain}
\ee
where $F_{k}$ is defined in (\ref{Fs}). 
The equations (\ref{Fconditionsagain}) ensure that the
Bethe vector corresponding to the singular solution 
(\ref{gensing}), namely
\be 
\lim_{\epsilon \rightarrow 0}\ |\lambda_{1} \,, \ldots \,, 
\lambda_{M} \rangle \,,
\label{gensingvec}
\ee
where $\lambda_{1}\,, \lambda_{2}$ are given by (\ref{regularized})
and $|\lambda_{1} \,, \ldots \,, \lambda_{M} \rangle$ is given by (\ref{vec}), is an
eigenvector of the transfer matrix. 

In other words, given a solution $\{  
\lambda_{3}\,, \ldots \,, \lambda_{M} \}$ of (\ref{BAE2}),
if the equations 
(\ref{Fconditionsagain}) {\it can} be satisfied, then they determine the 
parameter $c$ in (\ref{regularized}), and the corresponding Bethe vector (\ref{gensingvec}) is an
eigenvector of the transfer matrix. We call such a singular solution  
``physical''.
On the other hand, if the equations 
(\ref{Fconditionsagain}) {\it cannot} be satisfied, then -- despite the 
fact that the usual Bethe equations (\ref{BAE}), (\ref{BAE2}) are 
obeyed -- this solution cannot be used to construct an eigenvector of the transfer 
matrix. We call such a singular solution  ``unphysical''. Hence, 
according to the previous section, all singular solutions with odd
$N$ and $M=2$ are unphysical.  

Eqs. (\ref{Fconditionsagain}) can be simplified as follows. Using  
(\ref{regularized}), we find that these two equations imply
\be
c = 
-\frac{2}{i^{N+1}}\prod_{j=3}^{M}\frac{\lambda_{j}-\frac{3i}{2}}{\lambda_{j}+\frac{i}{2}} \,, \qquad
c = 
2i^{N+1}\prod_{j=3}^{M}\frac{\lambda_{j}+\frac{3i}{2}}{\lambda_{j}-\frac{i}{2}} \,,
\ee
respectively. These equations in turn imply the consistency condition
\be
\prod_{j=3}^{M}\left(\frac{\lambda_{j}-\frac{i}{2}}{\lambda_{j}+\frac{i}{2}}\right)
\left(\frac{\lambda_{j}-\frac{3i}{2}}{\lambda_{j}+\frac{3i}{2}}\right) = (-1)^{N} 
\,. 
\label{consistency1}
\ee 
By forming the product of all the Bethe equations (\ref{BAE2}), we obtain 
the relation
\be 
\prod_{k=3}^{M}\left(  \frac{\lambda_{k} + \frac{i}{2}}
{\lambda_{k} - \frac{i}{2}} \right)^{N-1} 
\left(\frac{\lambda_{k} - \frac{3i}{2}}
{\lambda_{k} + \frac{3i}{2}}\right) = 1 \,,
\ee
using which the consistency condition (\ref{consistency1}) takes the 
simple form
\be
\left[-\prod_{k=3}^{M}\left(  \frac{\lambda_{k} + \frac{i}{2}}
{\lambda_{k} - \frac{i}{2}} \right)\right]^{N} =1 \,.
\label{consistency2}
\ee

We remark that the condition (\ref{consistency2}) provides a practical
way to select from among the many singular solutions of the Bethe
equations (\ref{BAE2}) the physically relevant subset, which is
generally much smaller.  For example, for $N=6$ and $M=3$, the Bethe
equations (\ref{BAE}), (\ref{BAE2}) have 5 singular solutions, of
which only one is physical.  Similarly, for $N=8$ and $M=4$, we find
21 singular solutions, of which only 3 are physical.\footnote{The
number of singular states of the XXZ chain are estimated in
\cite{Noh:2000}.}

\section{Conclusion}\label{sec:conclusion}

We have seen that the ABA for the isotropic periodic Heisenberg chain
must be extended for solutions of the Bethe equations containing $\pm
\frac{i}{2}$.  Indeed, such singular solutions must be carefully regularized as
in (\ref{regularized}) or (\ref{regularized2}).  This regularization
involves a parameter that can be determined using a generalization of
the Bethe equations given by (\ref{Fconditions}), where $F_{k}$ is
defined in (\ref{Fs}).  These equations also provide a practical way
of determining which singular solutions correspond to eigenvectors of
the model.  In particular, the solution $\pm \frac{i}{2}$ must be
excluded for odd $N$.

It would be interesting to know whether the finite-$\epsilon$
corrections to the energy have any physical significance.  We expect
that our analysis can be extended to the anisotropic case.

\vspace{0.2in}
\noindent{\em Note Added}

After completing this work, we became aware of \cite{Avdeev:1985cx}, where similar 
results were obtained for the solution $\pm \frac{i}{2}$. However, our approach differs 
significantly from theirs.

\section*{Acknowledgments}
We thank Omar Foda, Vitaly Tarasov and Michael Wheeler for helpful 
correspondence, and Vladimir Korepin for reading a draft.
This work was supported in part by the National Science Foundation 
under Grants PHY-0854366 and PHY-1212337, and by a Cooper fellowship.

\appendix

\section{Appendix}

Here we fill in some details. It is convenient to define an 
unrenormalized Lax operator (as in \cite{Faddeev:1996iy}):
\be
\tilde L_{n a}(\lambda) = (\lambda - 
\frac{i}{2})\id_{n a} + i {\cal P}_{n a} \,,
\label{tildeLax}
\ee
and correspondingly
\be
\tilde T_{a}(\lambda) = \tilde L_{N a}(\lambda)  \cdots \tilde L_{1 a}(\lambda) = \left(
\begin{array}{cc}
    \tilde A(\lambda) & \tilde B(\lambda) \\
    \tilde C(\lambda) & \tilde D(\lambda) 
\end{array}    \right) \,.
\label{TT}
\ee 
Evidently,
\be
L_{n a}(\lambda) =  \frac{1}{(\lambda + \frac{i}{2})}\, \tilde L_{n 
a}(\lambda)\,, \qquad
T_{a}(\lambda) = \frac{1}{(\lambda + \frac{i}{2})^{N}}\, \tilde  
T_{a}(\lambda) \,.
\ee 
In particular, 
\be
B(\lambda) = \frac{1}{(\lambda + \frac{i}{2})^{N}}\, 
\tilde B(\lambda) \,.
\label{BtildeBreltn}
\ee 
Since $\tilde B(\pm \frac{i}{2})$ are finite, it 
follows that $B(\frac{i}{2})$ is also finite, and
\be
B(-\frac{i}{2} +\epsilon) \sim \frac{1}{\epsilon^{N}}  
\label{Bbehavior}
\ee
plus less singular terms.

The fact (\ref{Bbehavior}) suggests that $|\lambda_{1}^{naive} \,,  
\lambda_{2}^{naive} \rangle = B(\frac{i}{2} +\epsilon)\,  
B(-\frac{i}{2} +\epsilon) |0\rangle$ should be similarly divergent 
for $\epsilon \rightarrow 0$. However, we shall now argue that 
this vector is in fact finite. In view of (\ref{BtildeBreltn}), it 
suffices to show that\footnote{The result (\ref{claim}) implies, as 
already noted, that this vector is null in the limit $\epsilon \rightarrow 0$.}
\be
\tilde B(\frac{i}{2} +\epsilon)\,  
\tilde B(-\frac{i}{2} +\epsilon) |0\rangle \sim \epsilon^{N} \,.
\label{claim}
\ee
To this end, we proceed by induction. The behavior (\ref{claim}) can be easily verified 
explicitly for $N=4$ using {\tt Mathematica}. We observe from (\ref{TT}) that the 
monodromy matrices for $N-1$ and $N$ sites are related by
\be
\tilde  T_{a}^{(N)}(\lambda) = \tilde  L_{N a}(\lambda)\, \tilde  
T_{a}^{(N-1)}(\lambda) \,,
\ee
which implies that
\be
\left(
\begin{array}{cc}
    \tilde A^{(N)}(\lambda) & \tilde B^{(N)}(\lambda) \\
    \tilde C^{(N)}(\lambda) & \tilde D^{(N)}(\lambda) 
\end{array}    \right)  = 
\left(
\begin{array}{cc}
    \tilde a_{N}(\lambda) & \tilde b_{N}(\lambda) \\
    \tilde c_{N}(\lambda) & \tilde d_{N}(\lambda) 
\end{array}    \right)\,
\left(
\begin{array}{cc}
    \tilde A^{(N-1)}(\lambda) & \tilde B^{(N-1)}(\lambda) \\
    \tilde C^{(N-1)}(\lambda) & \tilde D^{(N-1)}(\lambda) 
\end{array}    \right)
\,,
\ee 
where 
\be
\tilde a_{N}(\lambda) &=&  \left(\begin{array}{cc}
\lambda+\frac{i}{2} & 0 \\ 
0  & \lambda-\frac{i}{2}
\end{array}    \right)\,, \qquad 
\tilde b_{N}(\lambda) = \left(\begin{array}{cc}
0 & 0 \\
i & 0 
\end{array}    \right)\,, \non \\
\tilde c_{N}(\lambda) &=& \left(\begin{array}{cc}
0 & i \\
0 & 0 
\end{array}    \right)\,,  \qquad \qquad \qquad
\tilde d_{N}(\lambda) = \left(\begin{array}{cc}
\lambda-\frac{i}{2} & 0 \\ 
0  & \lambda+\frac{i}{2}
\end{array}    \right) \,.
\ee
In particular,
\be
\tilde B^{(N)}(\lambda) = \tilde a_{N}(\lambda)\, \tilde 
B^{(N-1)}(\lambda) + \tilde b_{N}(\lambda)\, \tilde D^{(N-1)}(\lambda) \,.
\ee 
It follows that
\be
\tilde B^{(N)}(\lambda_{1})\,  
\tilde B^{(N)}(\lambda_{2}) |0\rangle^{(N)} &=&
\left[  \tilde a_{N}(\lambda_{1})\, \tilde 
B^{(N-1)}(\lambda_{1}) + \tilde b_{N}(\lambda_{1})\, \tilde 
D^{(N-1)}(\lambda_{1}) \right] \non \\
&\times &  \left[  \tilde a_{N}(\lambda_{2})\, \tilde 
B^{(N-1)}(\lambda_{2}) + \tilde b_{N}(\lambda_{2})\, 
\tilde D^{(N-1)}(\lambda_{2}) \right] 
|0\rangle^{(N-1)}  {1 \choose 0}_{N}  \non \\
&=& \Bigg[ \tilde a_{N}(\lambda_{1})\, \tilde a_{N}(\lambda_{2})\,
\tilde B^{(N-1)}(\lambda_{1})\, \tilde B^{(N-1)}(\lambda_{2}) \non \\
&& +\ \tilde a_{N}(\lambda_{1})\, \tilde b_{N}(\lambda_{2})\,
\tilde B^{(N-1)}(\lambda_{1})\, \tilde D^{(N-1)}(\lambda_{2}) \non \\
&& +\ \tilde b_{N}(\lambda_{1})\, \tilde a_{N}(\lambda_{2})\,
\tilde D^{(N-1)}(\lambda_{1})\, \tilde B^{(N-1)}(\lambda_{2})   \non \\
&&+\ \tilde b_{N}(\lambda_{1})\, \tilde b_{N}(\lambda_{2})\,
\tilde D^{(N-1)}(\lambda_{1})\, \tilde D^{(N-1)}(\lambda_{2}) \Bigg] 
|0\rangle^{(N-1)} {1 \choose 0}_{N} \label{prod}
\ee
for $\lambda_{1}\,, \lambda_{2}$ arbitrary.

We now set $\lambda_{1} = \lambda_{1}^{naive} = \frac{i}{2} + \epsilon$ 
and  $\lambda_{2} = \lambda_{2}^{naive} = -\frac{i}{2} + \epsilon$, 
and we consider the four terms on the RHS of (\ref{prod}), starting 
with the first: by the induction hypothesis,
\be
\tilde B^{(N-1)}(\lambda_{1})\, \tilde B^{(N-1)}(\lambda_{2}) 
|0\rangle^{(N-1)}  \sim \epsilon^{N-1} \,.
\ee
Moreover, it is easy to see that
\be
\tilde a_{N}(\lambda_{1})\, \tilde a_{N}(\lambda_{2}) {1 \choose 
0}_{N} \sim \epsilon \,.
\ee
Hence, the first term on the RHS of  (\ref{prod}) is of 
order $\epsilon^{N}$.

The fourth term on the RHS of (\ref{prod}) is zero because
\be
\tilde b_{N}(\lambda_{1})\, \tilde b_{N}(\lambda_{2}) {1 \choose 
0}_{N} = 0\,.
\ee

Using the exchange relation \cite{Faddeev:1996iy}
\be
\tilde D(\lambda_{1})\, \tilde B(\lambda_{2}) = 
\frac{\lambda_{1}-\lambda_{2}+i}{\lambda_{1}-\lambda_{2}}\tilde 
B(\lambda_{2})\, \tilde D(\lambda_{1}) - 
\frac{i}{\lambda_{1}-\lambda_{2}}\tilde 
B(\lambda_{1})\, \tilde D(\lambda_{2})
\ee
in the third term, we see that the second and third terms on the RHS 
of (\ref{prod}) combine to give
\be
\lefteqn{\Bigg\{\left[ \tilde a_{N}(\lambda_{1})\, \tilde b_{N}(\lambda_{2}) -
\tilde b_{N}(\lambda_{1})\, \tilde a_{N}(\lambda_{2}) \right] 
\tilde B^{(N-1)}(\lambda_{1})\, \tilde D^{(N-1)}(\lambda_{2})} \non \\
&&+\ 2\tilde b_{N}(\lambda_{1})\, \tilde a_{N}(\lambda_{2})
\tilde B^{(N-1)}(\lambda_{2})\, \tilde  D^{(N-1)}(\lambda_{1}) 
\Bigg\} |0\rangle^{(N-1)} {1 \choose 0}_{N} \label{combined}
\ee 
The first line of (\ref{combined}) gives a vanishing contribution because
\be
\left[ \tilde a_{N}(\lambda_{1})\, \tilde b_{N}(\lambda_{2}) -
\tilde b_{N}(\lambda_{1})\, \tilde a_{N}(\lambda_{2}) \right] {1 
\choose 0}_{N} =0 \,.
\ee 
The second line of (\ref{combined}) is of order $\epsilon^{N}$, since
\be
\tilde  D^{(N-1)}(\lambda_{1}) |0\rangle^{(N-1)} \sim \epsilon^{N-1} \,,
\ee 
and 
\be
\tilde b_{N}(\lambda_{1})\, \tilde a_{N}(\lambda_{2}) {1 \choose 0}_{N} \sim \epsilon \,.
\ee 
In short, we have shown that
\be
\tilde B^{(N)}(\lambda_{1})\,  
\tilde B^{(N)}(\lambda_{2}) |0\rangle^{(N)} \sim \epsilon^{N} \,,
\ee
which concludes the inductive proof of our claim (\ref{claim}).

% \bibliographystyle{utphys}
% \bibliography{intads_rev,refs}

\begin{thebibliography}{10}

\bibitem{Faddeev:1996iy}
L.~Faddeev, ``{How algebraic Bethe ansatz works for integrable model},''
\href{http://arxiv.org/abs/hep-th/9605187}{{\ttfamily arXiv:hep-th/9605187
  [hep-th]}}.
%%CITATION = HEP-TH/9605187;%%.

\bibitem{Korepin:1993}
V.~Korepin, N.~Bogoliubov, and A.~Izergin, {\em {Quantum Inverse Scattering
  Method and Correlation Functions}}.
\newblock Cambridge University Press, 1993.

\bibitem{Siddharthan:1998}
R.~Siddharthan, ``{Singularities in the Bethe solution of the XXX and XXZ
  Heisenberg spin chains},''
  \href{http://arxiv.org/abs/cond-mat/9804210}{{\ttfamily
  arXiv:cond-mat/9804210 [cond-mat]}}.

\bibitem{Wal:1999}
A.~Wal, T.~Lulek, B.~Lulek, and E.~Kozak, ``{The Heisenberg magnetic ring with
  6 nodes: exact diagonalization, Bethe ansatz and string configurations},''
  {\em Int. J. Mod. Phys. B} {\bfseries 13} (1999) 3307.

\bibitem{Essler:1991wd}
F.~H. Essler, V.~E. Korepin, and K.~Schoutens, ``{Fine structure of the Bethe
  ansatz for the spin 1/2 Heisenberg XXX model},''
{\em J.Phys.} {\bfseries A25} (1992) 4115--4126.
%%CITATION = ITP-SB-91-48 ETC.;%%.

\bibitem{Beisert:2003xu}
N.~Beisert, J.~A. Minahan, M.~Staudacher, and K.~Zarembo, ``{Stringing spins
  and spinning strings},'' {\em JHEP} {\bfseries 09} (2003) 010,
\href{http://arxiv.org/abs/hep-th/0306139}{{\ttfamily arXiv:hep-th/0306139}}.
%%CITATION = HEP-TH/0306139;%%.

\bibitem{Beisert:2004hm}
N.~Beisert, V.~Dippel, and M.~Staudacher, ``{A novel long range spin chain and
  planar $\mathcal{N}$ = 4 super Yang-Mills},'' {\em JHEP} {\bfseries 07}
  (2004) 075,
\href{http://arxiv.org/abs/hep-th/0405001}{{\ttfamily arXiv:hep-th/0405001}}.
%%CITATION = HEP-TH/0405001;%%.

\bibitem{Goetze:2010}
W.~Goetze, U.~Karahasanovic, and F.~Essler, ``{Low-Temperature Dynamical
  Structure Factor of the Two-Leg Spin-1/2 Heisenberg Ladder},'' {\em Phys.
  Rev. B} {\bfseries 82} (2010) 104417,
  \href{http://arxiv.org/abs/1005.0492}{{\ttfamily arXiv:1005.0492
  [cond-mat]}}.

\bibitem{Arutyunov:2012tx}
G.~Arutyunov, S.~Frolov, and A.~Sfondrini, ``{Exceptional Operators in N=4
  super Yang-Mills},'' \href{http://dx.doi.org/10.1007/JHEP09(2012)006}{{\em
  JHEP} {\bfseries 1209} (2012) 006},
\href{http://arxiv.org/abs/1205.6660}{{\ttfamily arXiv:1205.6660 [hep-th]}}.
%%CITATION = ARXIV:1205.6660;%%.

\bibitem{Mukhin:2009}
E.~Mukhin, V.~Tarasov, and A.~Varchenko, ``{Bethe algebra of homogeneous XXX
  Heisenberg model has simple spectrum},'' {\em Commun. Math. Phys.} {\bfseries
  288} (2009) 1--42, \href{http://arxiv.org/abs/0706.0688}{{\ttfamily
  arXiv:0706.0688 [math]}}.

\bibitem{Vladimirov:1986}
A.~Vladimirov, ``{Proof of the invariance of the Bethe-ansatz solutions under
  complex conjugation},'' {\em Theor. Math. Phys.} {\bfseries 66} (1986) 102.

\bibitem{Noh:2000}
J.~D. Noh, D.-S. Lee, and D.~Kim, ``{Origin of the singular Bethe ansatz
  solutions for the Heisenberg XXZ spin chain},'' {\em Physica A} {\bfseries
  287} (2000) 167.

\bibitem{Avdeev:1985cx}
L.~Avdeev and A.~Vladimirov, ``{Exceptional solutions of the Bethe ansatz
  equations},''
\href{http://dx.doi.org/10.1007/BF01037864}{{\em Theor. Math. Phys.} {\bfseries
  69} (1987) 1071}.
%%CITATION = TMPHA,69,1071;%%.

\end{thebibliography}
    
\providecommand{\href}[2]{#2}\begingroup\raggedright\endgroup

\end{document}